\documentclass{article}
\usepackage[utf8]{inputenc}

\title {Reinforcement Learning for Electricity Network Operation \\
\large Learning to Run a Power Network 2020 Challenge - White Paper}
\author{Adrian Kelly (Electric Power Research Institute - EPRI) \\ Aidan O'Sullivan, Patrick de Mars (University College London - UCL), \\Antoine Marot (RTE - Réseau de Transport d'Electricité)}
\date{March 2020}

\usepackage[numbers]{natbib}
\usepackage{graphicx}
\usepackage{caption}
\usepackage{subcaption}
\usepackage{amsmath, amssymb}

\begin{document}

\maketitle

\section{Introduction to the Challenge}
The goal of this challenge is to test the potential of Reinforcement Learning (RL) to control electrical power transmission, in the most cost-effective manner, while keeping people and equipment safe from harm. Solving this challenge may have very positive impacts on society, as governments move to decarbonize the electricity sector and to electrify other sectors, to help reach IPCC climate goals. 
Existing software, computational methods and optimal powerflow solvers are not adequate for real-time network operations on short temporal horizons in a reasonable computational time. With recent changes in electricity generation and consumption patterns, system operation is moving to become more of a stochastic rather than a deterministic control problem.  In order to overcome these complexities, new computational methods are required. The intention of this challenge is to explore RL as a solution method for electricity network control. There may be under-utilized, cost-effective flexibility in the power network that RL techniques can identify and capitalize on, that human operators and traditional solution techniques are unaware of or unaccustomed to. 

An RL agent that can act in conjunction, or in parallel with human network operators, will optimize grid security and reliability, allowing more renewable resources to be connected while minimizing the cost and maintaining supply to customers, and preventing damage to electrical equipment.

Another aim of the project is to broaden the audience for the problem of electricity network control and to foster collaboration between experts in both the power systems community and the wider RL/ML community.   

 \subsection{The Importance of Electricity}
  Electricity is an underrated, and under-appreciated foundational element of modern life. With the invention of electricity networks in the early 20\textsuperscript{th} century, electricity can be considered one of the most important factors for the growth of economies and the functioning of societies around the world. It is a vital resource in daily living, industry, agriculture and public transportation. The electrification of developing countries and rural communities has helped to lift billions of people out of poverty.  Blackouts of cities or countries, when they occur, are catastrophic events, which can cause chaos, and have a major impact on modern life. Recent high profile blackouts in Great Britain, Argentina, New York and Australia show the importance electricity has in society and the disruption it can cause when it is disconnected.  Electricity network operators, across the world, must ensure a constant, reliable, secure, safe, cost-effective and sustainable supply of electricity is maintained, while preventing blackouts. 
 
\begin{figure}[h!]
     \centering
     \includegraphics[width=3.5in]{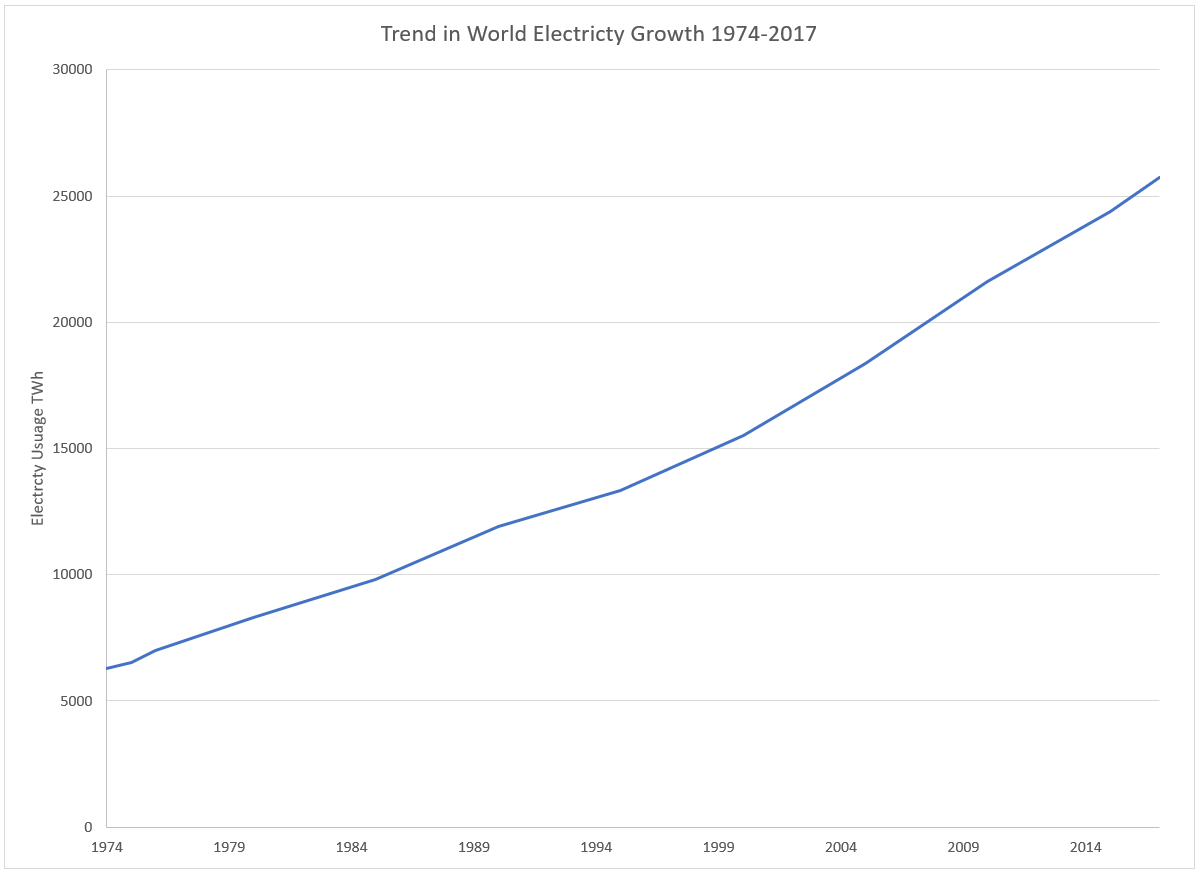}
     \caption{World electricity usage growth by sector from 1974-2017. Source: IEA (2019), "Electricity Information 2019", IEA, Paris https://www.iea.org/reports/electricity-information-2019 \cite{IEA2019}.}
     \label{fig:Netelectricityuse}
 \end{figure}

 To contextualize, an electric kettle uses abut 1500 Watts (W) of \textit{power} when switched on. If switched on for 1 hour it would use 1.5 kilowatt-hours (kWh) of \textit{energy}. In 2018 the average U.S. home used approximately 11 Megawatt-hours (MWh) of energy in the year, or 11,000,000 watt-hours \footnote{ Source: US Energy Information Administration See: https://www.eia.gov/tools/faqs/faq.php?id=97\&t=3}. In 2017 the total energy produced by electricity in the entire world was approximately 25,721 TWh or 25,721 trillion watt-hours \footnote{Source: International Energy Agency See: https://www.iea.org/data-and-statistics}. 
 Controlling electricity - from generation sources to end-users' kettles - is an extremely complex task. Meeting worldwide electricity demand, while electrifying sectors of society and developing countries, with relatively few blackouts, is why the electricity network has been described as the greatest human-made machine ever built. In 1974 total global electricity use was approximately 5 TWh. Electricity use has increased four-fold in less than 40 years. See Figure \ref{fig:Netelectricityuse} for a chart showing the growth of electricity usage since 1974 broken down by end-use application. In recent years the difficulty maintaining this level of security has increased. This challenge aims to help network operators to control the networks and to guide the decision making of network operators in real-time operations and in operational planning. 
 
\section{Introduction to Electricity and Power Networks for the Machine Learning Community}
 In this section, there is a concise introduction to electricity concepts as they relate to the challenge with the aim of opening up the challenge up to the broader ML/RL community. 
 \\ \textbf{Note:} There will be a significant simplification of concepts and terms for the benefit of brevity and to emphasize the most important aspects to non-expert readers. For a more comprehensive exposition of the theory and principles of electricity and power systems please refer to \cite{grigsby2012} and \cite{vonmeier2006}.  
 
 \subsection{The Basic Physics of Electricity}
 \label{sec:elec_physics}
 Electricity is a form of energy involving the \textit{excitement} of electrons in metallic elements. This phenomenon is not unique to electricity networks, but is also critical to life itself, as our bodies are dependent on electrical pulses for the functioning of our muscles, cells and nerves. In order to develop an understanding of electricity, it is necessary to introduce the fundamental dimension of physical measurement \textit{electric charge}. Charge is a property of matter arising from atomic structure which is made up of protons (positively charged), electrons (negatively charged) and neutrons (neutral). It is measured in coulombs (C), a charge equal to that of $6.25 \times 10^{18}$ protons. \\
 Charge induces a force with opposite charges attracting and the same charges repelling. This force creates the ability to produce work and the electric potential or \textit{voltage}, which is the potential energy possessed by a charge at a location relative to a reference location. It is defined between two points and measured in Volts, denoted with the symbol V. An electric  \textit{current}  is a \textit{flow} of charge through a  material,  measured in Coulombs per second or Amperes (A) and denoted with the symbol I. (While described as a flow, nothing new is actually created or moves, current flow in the electricity context means the excitement of electrons in metallic conductors to induce a voltage and produce charge along the metallic conductor).
\\
\\The electrical power is given as the product of the voltage and the current. 
    \begin{equation}
     P = VI
    \label{eq:1}
    \end{equation}
 
Power is measured in Watts, denoted by the symbol W, see section 1.1 for how this is related to everyday electricity usage. In order to try to simplify these electrical concepts, an analogy with a physical water system is often used, while not quiet directly analogous, the current is similar to the flow of water in a pipe, say in litres per second. Voltage would be analogous to a height difference, say between a water reservoir and the downhill end of the pipe, or a pressure difference. Intuitively, voltage is a measure of `how badly the material wants to get there' and current is a measure of `how much material is actually going'. Power would be analogously produced by the force of water spinning a hypothetical turbine that may rotate a wheel. Intuitively these phenomena are related, increasing the voltage or current in a system increases the power produced. Electrically, this relationship is captured by Ohm's law:
 
 \begin{equation}
     V = IR
\label{eq:2}
 \end{equation}
 
A new variable is introduced here - $R$ - which is the \textit{resistance} of the material the current is \textit{flowing} through, analogous to the size of the water-pipe. A smaller pipe makes it harder for large flows and it is the same with current - highly conductive materials allowing current to \textit{flow} easily and poorly conductive materials (called insulators) preventing current from \textit{flowing}. Whenever an electric current exists in a material with resistance it will create heat. The amount of heating is related to the power P and combining equations \eqref{eq:1} and \eqref{eq:2} gives:
 
 \begin{equation}
     P = I^2R
\label{eq:3}
 \end{equation}
 Heating can be desirable. Heating a resistive element is how an electric kettle or heater works. It can also be undesirable - as is the case of power lines - where the heat is energy lost and causes thermal expansion of the conductor making them \textit{sag}, or fall close to the ground or to people or buildings. In extreme cases, such as a fault condition, thermal heating can melt the wires. As we see from Equation \eqref{eq:3} the amount of heating is proportional to the square of the current, so increasing the current has a large effect on the resistive losses. It is for this reason that when electricity is transported over long distances, it is done at high voltages. Based on Equation \eqref{eq:2}, assuming that the resistance of the line remains constant, to transport the same amount of power, resistive losses are reduced by increasing the voltage and lowering the current. This is $the$ fundamental concept of electricity transmission.
 
 \begin{figure}[h!]
     \centering
     \includegraphics[width=2.5in]{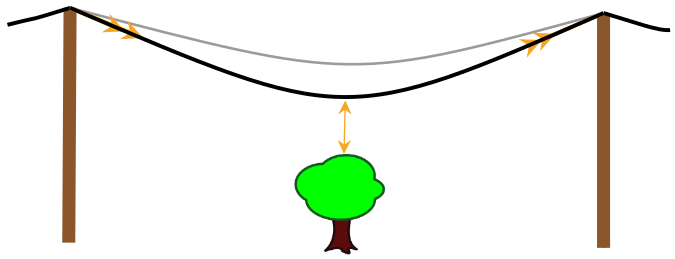}
     \includegraphics[width=2.5in]{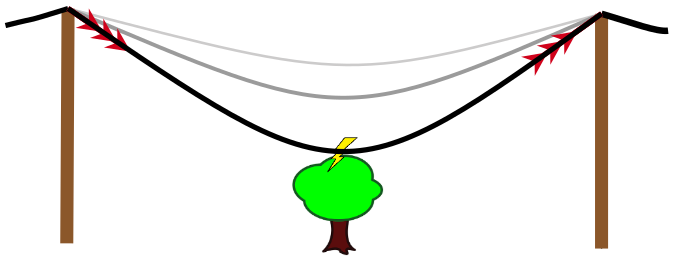}
     \caption{An example of the dangers of overheating power-lines, by transporting too much current, the metallic conductor heats and sags close to the ground causing a flash over to ground and endangering human life.}
     \label{fig:poweline_tense}
 \end{figure}
 
\begin{figure}[h!]
     \centering
     \includegraphics[width=2.5in]{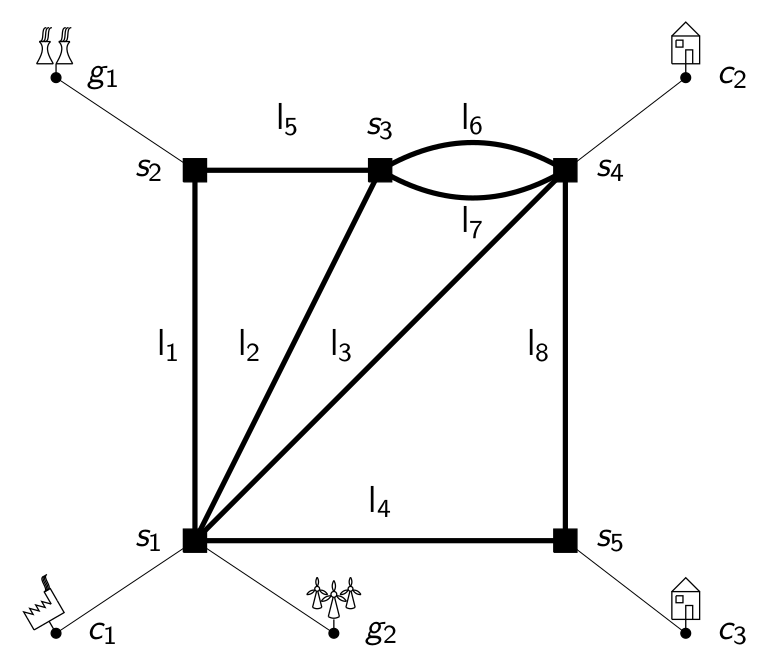}
     \caption{A simple electricity network, showing the circuit nature of a power network, the currents I flowing in the lines and the \textit{interconnectedness} between generators denoted g, customer loads denoted c and substation nodes denoted s.}
     \label{fig:grid}
 \end{figure}

 In order to produce a sustained flow of current, the voltage must be maintained on the conductor. This is achieved by providing a pathway to recycle charge to its origin and a mechanism, called an electromotive force (emf), that compels the charge to return to its original potential. Such a setup constitutes an electric circuit. Again, to oversimplify by relating back to the water analogy - if there is an open pipe in the circuit water will run out. Likewise, if there is a break in an electric circuit, current will not flow but voltage will still be present on the conductor. Simple electric circuits are often described in terms of their constituent components; voltage sources, conductors and resistances. Complex power networks can be described in terms of generation sources, network lines and loads. A simple electrical power network analogous to a simple electric circuit is shown in Figure \ref{fig:grid}. 
 
Circuit analysis is the goal of estimating the parameters in a circuit given a combination of the voltages, currents and resistances and the fundamental Equations \eqref{eq:1}, \eqref{eq:2} and \eqref{eq:3}. The more complex the circuit or network, the more complex the analysis will be. Within a circuit, a series of laws known as Kirchhoff's law also help us in the analysis: 

\begin{itemize}
    \item Kirchhoff's voltage law: voltage around any closed loop sums to zero
    \item Kirchhoff's current law: current entering and exiting any node sums to zero
\end{itemize}

These principles can be applied at the micro-level to simple circuits, such as plugging in an electric kettle where the element is the resistor or load, the mains outlet is the voltage source and current is proportional to the voltage and resistance of the circuit. Voltage is maintained throughout the circuit when it is plugged in and current \textit{flows} from the plug outlet through the wire, into the heating element and back to the plug outlet, completing the circuit. 
\\ These concepts can also be applied at the macro level, where a house or town could be considered the load and a nuclear power station could be considered the voltage and current source, which is interconnected to the load by power lines. The electricity network is one large circuit, which is constantly satisfying these laws.

\subsection{Power Generation, Transmission and Consumption}

Now that the theoretical background on the physics of electricity has been outlined at a high level, the discussion can move to how power is produced and transported and consumed in national power networks (also referred to as grids). 
\subsubsection{Power Generation}
\begin{figure}[h!]
    \centering
    \includegraphics[width=2.5in]{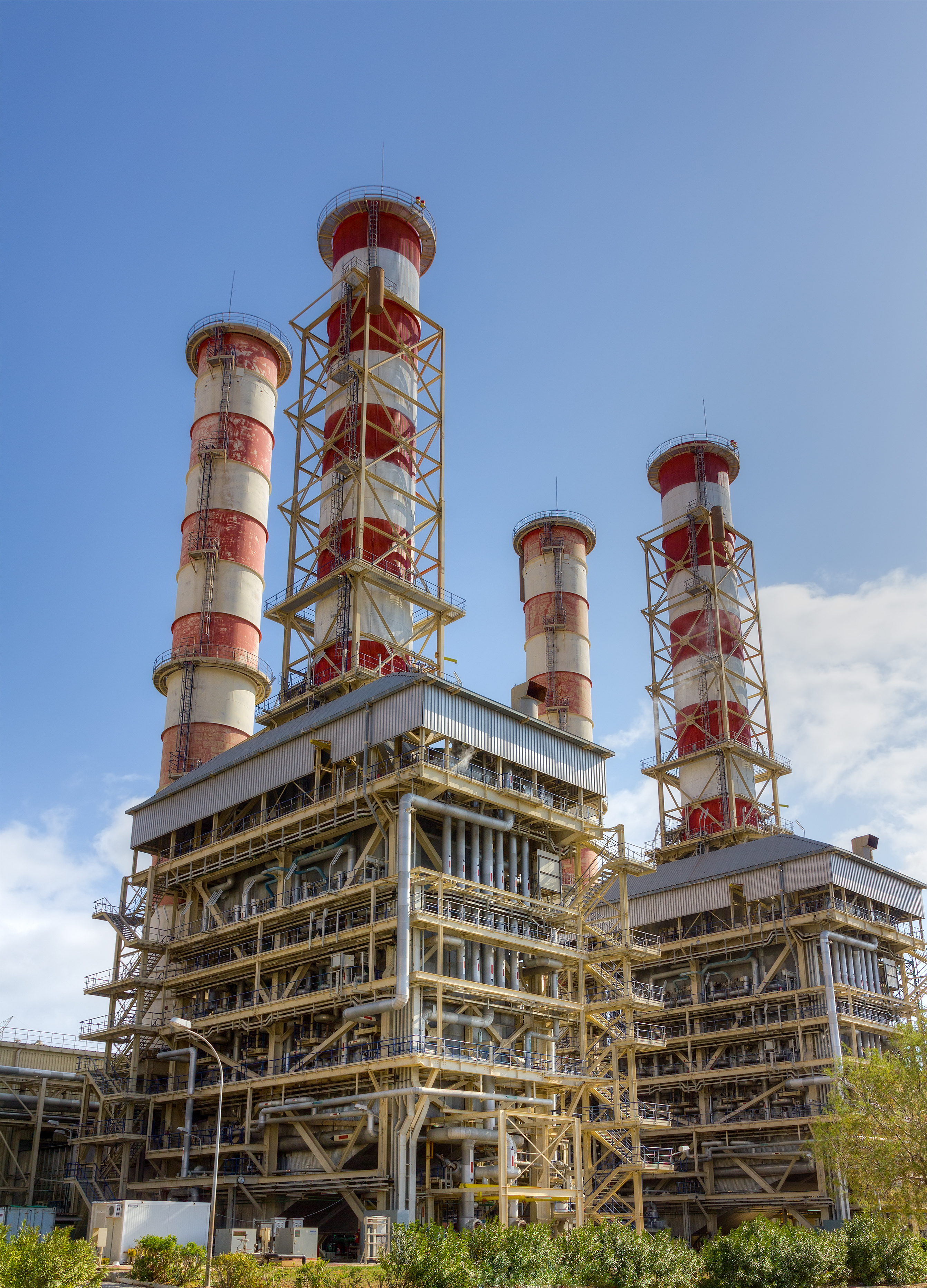}
    \includegraphics[width=2.5in]{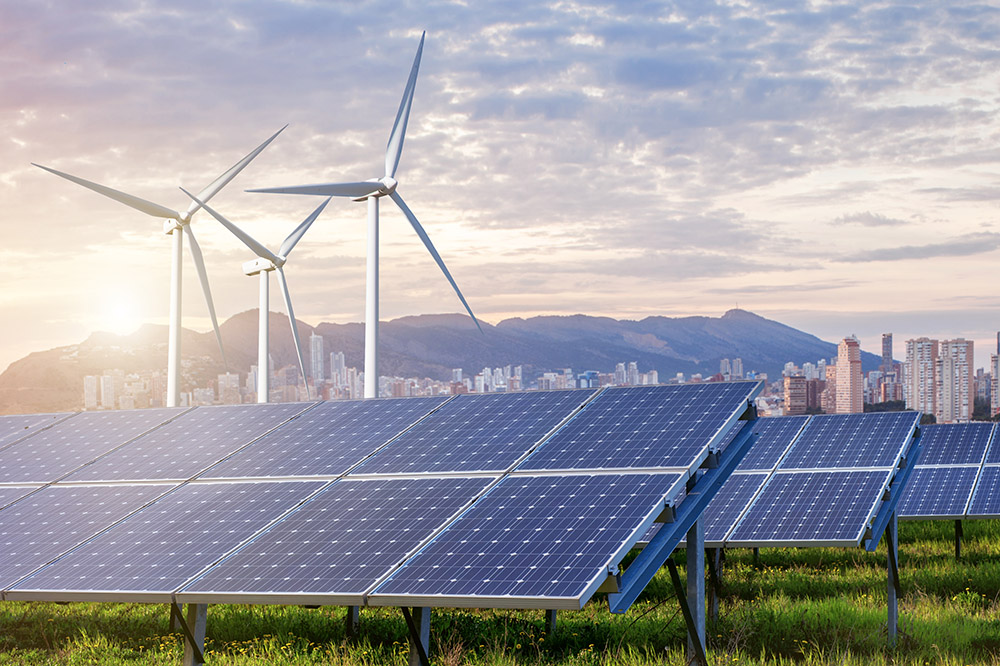}
    \caption{Examples of generation sources, a thermal power generator station and at the bottom, solar panels and wind turbines.} 
    \label{fig:Gen}
\end{figure}

On a network, power is provided from multiple different technologies using different fuels which are all referred to as generators. They can be considered as sources in the power network. Traditionally power was generated by large, thermal units burning fossil fuels such as coal, oil and gas. These large generators were co-located with load, in cities and towns, thereby reducing the distance that power needed to be transmitted. In recent years due to the shifts in policy to decarbonize society and in the liberalization of electricity markets,  generation sources have shifted to renewable, unpredictable, weather-based sources such as wind and solar.  These sources are often installed in geographically diverse and less populated areas and produce power far away from load centres. Hydro and nuclear power stations, while not new, are carbon-free and are also located relatively far away from load centres. The network needs to be planned and operated in different ways, to incorporate geographically disperse, variable generation sources and ensure that power is efficiently transmitted to the load centres at all times. See Figure \ref{fig:Gen} for pictorial examples of generation sources.

\subsubsection{Transmission}
\begin{figure}[h!]
    \centering
    \includegraphics[width=5in]{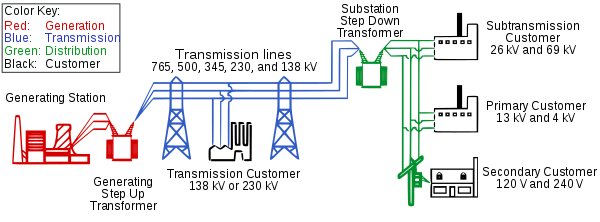}
    \caption{Graphical illustration of power networks from generation to transmission to consumption. Source: Wikipedia.} 
    \label{fig:Trans}
\end{figure}

When power networks first came into existence there was a great controversy over what the best approach to the transmission of electricity might be. This became known as the war of the currents and was played out with the main protagonists; Thomas Edison who supported DC transmission (Direct Current) versus George Westinghouse and Nikola Tesla who supported AC transmission (Alternating Current). AC power systems were technically superior and emerged the victor from this battle and all power networks in the world work on AC principles. Both AC and DC allow transmission at high voltage (thus reducing current), but the main reason for using AC (Alternating Current) in power networks is that it allows the raising and lowering of voltages using power transformers. Being able to increase the voltage allows us to transmit electricity greater distances due to the lower resistive heating losses, discussed in Section \ref{sec:elec_physics}. However, the voltage tends to vary based on the load on the network at particular times. Network operators must maintain the voltage at its nominal levels at all times on the transmission network. The transmission network is made up of transmission equipment, most importantly, overhead lines and underground cables which interconnect with substations. The substations contain the connections to the generation sources and the transformers to step up and step down the voltage to distribution systems. See Figure \ref{fig:Trans} for a simple schematic of the transmission grid. 
DC is a simpler linear system and is easier to understand, model and simulate, given there are no time-varying aspects. AC is more difficult as it introduces non-linearities based on sinusoidal aspects of voltage and current generation, three-phase transmission and mathematics in the complex domain. The RL challenge will run on an AC powerflow, but the important aspects of AC powerflow can be approximated by the linear equations of DC powerflow, with not much of a loss of accuracy of the challenge objective. Understanding DC powerflow is a good starting point to understanding power network control, and it is not necessary to have a deep understanding of AC powerflow to participate in the RL challenge.  

\subsubsection{Power Consumption}
When power is transmitted to substations, it is at too high a voltage for consumers such as homes and businesses to use. Transformers must be used to step-down the voltage to a low enough level for connection to distribution systems. Power is transmitted at between 100,000 - 760,000 volts. The voltage in homes is 220 Volts in Europe and 120 Volts in North America. From the power network viewpoint, power consumption is aggregated as load - calculated at the Megawatt level, and the power network operator's role generally ends when power is delivered to the step-down transformer. 

\subsection{Network Operation}
\subsubsection{How a Network Operates}
The power network is operated by ensuring the three primary constraints are met at all times, in all areas of the network.

\begin{itemize}
    \item Thermal limits of transmission equipment are not breached (measured in current with units of Amperes (A) or power with units MegaWatts (MW)).
    \item Voltage maintained within a defined range (measured in voltage, units of Volts (V)).
    \item Generation and load balanced at all times (measured in power, units of Megawatts (MW). The balance between load and generation is approximated by frequency measured in Hertz (Hz).
\end{itemize}

\subsubsection{Thermal Limits of Transmission Equipment}
Power will flow from source to load, around the network based on the resistance of the lines in the network. A transmission line has an upper limit to the amount of power that can flow through it before it will fail in service. This limit is given by the thermal properties of the metallic materials, usually copper or aluminium and also cooling and heating due to weather conditions (such as wind, irradiance and ambient temperature). If too much power is forced through the equipment for a long period, and the thermal limits are breached, the equipment is likely to fail in service and be disconnected. In reality, this means overhead lines sag closer to the ground, and may cause flashover as shown in Fig. \ref{fig:poweline_tense} (the cause of the 2003 blackout in North America) or very expensive equipment such as transformers or cables will be damaged and explode. It is better to disconnect the line than let it sag close to the ground. When the line is disconnected, the same amount of power is still present on the network, but one link has been removed. This means that the power will reroute itself to the new most desirable path based on the resistance, but this rerouting may result in another line or lines being overloaded. The challenge of network operation (and the basis of the RL challenge) is to route the power around the network in the most efficient manner, while avoiding overloads and cascading effects. 
\subsubsection{Voltage}
One of the key concepts of transmission is to step up and down the voltages on the network to optimize powerflow. Because of this, the transmission network substation nodes must be maintained at, or near the nominal voltage of the transmission equipment. If the voltage is not maintained, it will collapse and cause the system to blackout. Voltage collapse has been the cause of major network disturbances around the world. While not discussed in detail here, voltage is maintained by the reactive power balance on the network. Generation sources and power electronic devices produce reactive power and loads (such as motors) absorb reactive power. Transmission equipment like lines and cables also produce and consume reactive power for effective transmission of active power. The voltage of the system can vary around nominal values, it can be within 90 \% - 110 \% of its nominal at any time. For the RL challenge; voltage will not have to be optimized by the agent and will be controlled on the environment side, owing to the non-linear complexity that would have to be introduced into the problem. Voltage control may be introduced into future versions of the challenge.   

\subsubsection{Generation Load Balance and Frequency}
On power networks, the third constraint to consider is that generation produced by sources and the power consumed by loads must be equal at all times. The proxy measure for how balanced the system is the system frequency. The nominal frequency in Europe is 50 Hz and in North America, it is 60 Hz. If there is too little generation relative to demand, the frequency will drop below nominal. If there is too much generation relative to demand the frequency will rise above nominal. If the frequency deviates too far beyond nominal (approximately 2 Hz) it will likely collapse the entire grid. Generation and demand should be controlled to maintain the delicate balance at all times between generation and load and to maintain a balanced system frequency. For the most part, this means dispatching generation sources to meet the load demanded by end-users. For the RL challenge, the generation load balance will not have to be optimized and will be managed on the environment side of the model. Balancing of generation and load may be introduced into future versions of the challenge. However, dispatching generation sources is a useful action for power network control that can be utilized to alleviate line overloads in the RL challenge. 

\subsubsection{Network Impedance}
Returning to the concepts of circuit theory and the equations defined in Section 2, the pattern of powerflow around power network is governed by the resistance of the interconnecting lines. All equipment has a resistance, as well as a thermal limit as described above. Power cannot be created from a specific source and directly dispatched to a specific load. Power flows relative to the resistance of the lines connected between the generation sources and the loads, and it tends to take the path of least resistance. The lower the resistance, the more power will flow through the element. The line resistances can be considered to be constant and cannot be altered by operators in real-time and so is unlike thermal limits as it does not vary. However, the flows on the network can be altered by changing the topology of the grid by control actions such as by switching lines in and out. 

\subsubsection{Network States and Control}

\begin{figure}[h!]
     \centering
     \includegraphics[width=5in]{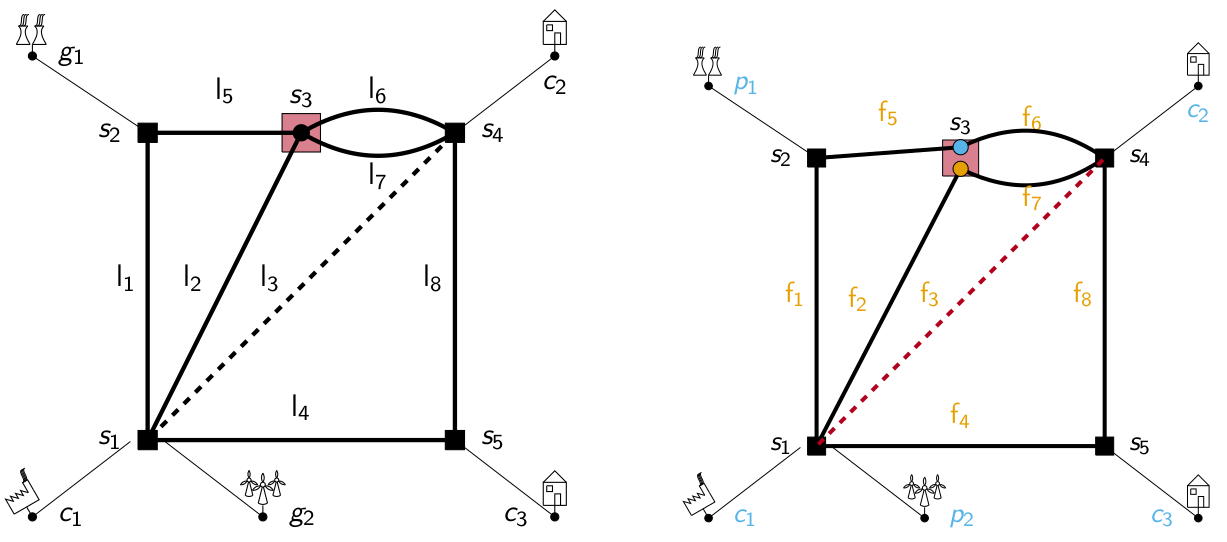}
     \caption{A simplistic network on the left, showing a simple network state, with no direct link between S1 and S4 since the dotted line indicates the line is switched out. The figure on the right shows a splitting action at substation node S3 which now creates an indirect link between node S1 and S4.}
     \label{fig:grid_2}
 \end{figure}

The network can be considered to be: 
\begin{itemize}
    \item Interconnected 
    \item Switchable 
    \item Static 
\end{itemize} 
Interconnected means it is meshed and not radial. This \textit{interconnectedness} allows a large amount of potential network states. 
Switchable means that most network elements have two states, i.e. lines can be set out-of-service or in service and substations can be split or coupled.
Static means that for network operators, new interconnected elements cannot be created in real-time. For example in Figure \ref{fig:grid_2} in the operations timeframe, a new line cannot be created between S2 to S5. But, with a limited amount of switching controls an indirect link can be created in real-time. 

\subsection{Network Security}
Secure operation of power networks is required under normal operating state as well as in contingency states. The following requirements must be met: 
\begin{itemize}
    \item In the normal operating state, the power flows on equipment, voltage and frequency are within pre-defined limits in real-time. 
    \item In the contingency state the power flows on equipment, voltage and frequency are within pre-defined limits after the loss of any single element on the network. 
\end{itemize}
The network must be operated to be secure in the “normal state”, i.e. the thermal limits on equipment must not be breached, the voltage must be within range and generation and load must be balanced. In the context of the RL challenge, the network must be operated such that no thermal limits on any lines are breached. The network must also be secure for any contingency, usually the loss of any element on the network (a generator, load, transmission element). Loss of elements can be anticipated (scheduled outages of equipment) or unanticipated (faults for lightning, wind, spontaneous equipment failure). Cascading failures must be avoided to prevent blackouts, this is the game over state in the RL challenge. The removal of equipment by protection equipment is in the millisecond time domain and is not considered in this challenge. 
Constraints on the system, such as line thermal overloads are alleviated in real-time by network operators using a range of remedial actions, from least to most costly as follows:
\begin{itemize}
    \item Switching lines on the network in or out \\
    \item Splitting or coupling busbars at substations together. This means a node can be split into two elements or connected together as a single element \\
    \item Redispatch generation to increase or reduce flows on lines \\
    \item Disconnecting load 
\end{itemize}
From a cost perspective, the disconnection of load should be avoided due to the disruption to society, business and daily life. Redispatching generation can also be expensive. The electricity is managed by a market, based on the cost per unit of energy supplied. If the network operators need to redispatch expensive generation, this can be sub-optimal from a market perspective and cause increased costs to customers. 
\\
To provide operational flexibility, substations are usually designed so that they can be separated into two or more constituent parts. Coupling a substation can serve to reroute power in a network and is an option to alleviate line overloads. Switching lines and coupling busbars at substations are the least costly option to alleviate thermal overloads on the network. There is considerable operational flexibility that is under-utilized on power networks that can be released by switching actions and topology changes. This network flexibility is easy to implement and the least costly option. One of the goals of the RL challenge is to explore the range of switching options available and to utilize topology changes to control power on the network. 

\subsubsection{Temporal Constraints}
There are also temporal elements to network operation, so network operation can be considered a time-domain challenge. The lines can be overloaded for short periods of time, while a solution to the issue is found. It is rare that an instantaneous overload will cause an instantaneous disconnection.  The load and renewable generation constantly change and the system must be secure at all times. The system must be operated so that load and generation are always in balance. In winter the peak load in the day might be at 6 PM to align with lighting, oven and heating load but some generation sources are ineffectual for this, in winter the sun may not shine at 6 PM, so solar generation is effectively useless at peak demand in winter. Large generation units cannot instantaneously connect and are dependent on the heat state of the metallic materials. Outages of equipment are scheduled, such as outages for maintenance or replacement or unscheduled such as spontaneous disconnections from lightning or other faults. 

\subsection{The Role of a Network Operator}
The transmission network is controlled from a control centre, with remote observability from this centre to all transmission network elements. The network operators can control most network elements such as lines and substations via remote control command.  The control centre also has visibility of all the generation sources and all the loads. Generation is controlled by the control centre operator sending dispatch instructions to change the outputs. Some loads can be controlled by the control centre, but, in general, the distribution system operators control switching of the load. For small to medium-sized countries, usually, there is one control centre with responsibility for network control but for larger countries like the USA, Canada there are multiple control centres that control the network at a state or regional level on a functional basis. These control centres coordinate their activities with their neighbouring control centres. 
\\
\\
The network operator's role is to monitor the electricity network 24 hours per day, 365 days per year. The operator must keep the network within its thermal limits, its frequency ranges and voltage ranges for  \textit{normal operation} and \textit{contingency state operation} as described above. 
For normal operation, the operator has a range of actions at their disposal to manage the network within its constraints, such as switching, generator dispatch and load disconnection. 
For the contingency state operation, the operator must act ahead of time to mitigate contingencies that may occur for the unexpected loss of any single element, using the prescribed range of actions. 
The operator must also plan the network operation for the loss of any element for a scheduled outage for maintenance. The network must operate securely for the entirety of the planned outage, not just for the moment the outage is taken. 
The operator must plan for and manage the network within its limits at the system peak, i.e. the largest load demand of the day, the generation must be managed so that the generation load balance (measured by the frequency) is maintained at the peak of the day. 
\subsection{How Can RL Benefit Network Operators}
Today on power networks around the world, the flexibility that topological changes to the network offers is an underexploited and low-cost option to maintaining network security. The issue is that the range of options is impossible to simulate in real-time and in the operations planning time frame, with the existing operator model and simulation toolkit. Typically operators will revert to their mental model and past experience for solutions to network problems when they arise. This worked in the past with a static network and predictable system operations. The new network operates differently and contains unpredictable resources, which can dramatically alter the flow patterns in the network. The deterministic study tools will no longer be adequate and new solutions for controlling the network, such as RL, will be required. 

\section{Introduction to Reinforcement Learning for Power Systems Community}
In this section, we give a brief overview of the field of reinforcement learning (RL) and how it applies to the Learning to Run Power Network challenge. 
 \\ \textbf{Note:} there will be a significant simplification of concepts and terms for the benefit of brevity and to emphasize the most important RL aspects for power system experts to understand. For a more comprehensive introduction to reinforcement learning please refer to \cite{Sutton2017}.
 \\
 \\
Machine learning is often characterised as consisting of three different branches: supervised learning (learning from labelled data), unsupervised learning (finding patterns in unlabelled data) and reinforcement learning, where an agent interacts with an environment to earn a reward. RL is analogous in many ways to how humans learn, through interaction with the real world through trial and error and feedback. Positive rewards encourage (or reinforce) good actions, while poor actions are penalised. RL provides a means of finding new strategies that improve on existing performance for a wide range of tasks. RL has had great success in achieving superhuman performance in games-playing, beating expert human players at some of the most complex and challenging games like Go \cite{SilverHuangEtAl16nature, Silver2017}. RL has also been used for learning the neuromuscular movement of a human-like agent in the previous NeurIPS challenge Learning To Run \cite{DBLP:journals/corr/abs-1804-00361}, which inspired this challenge. 

\subsection{Reinforcement Learning Problem Formulation}

\begin{figure}[h!]
    \centering
    \includegraphics[width=3in]{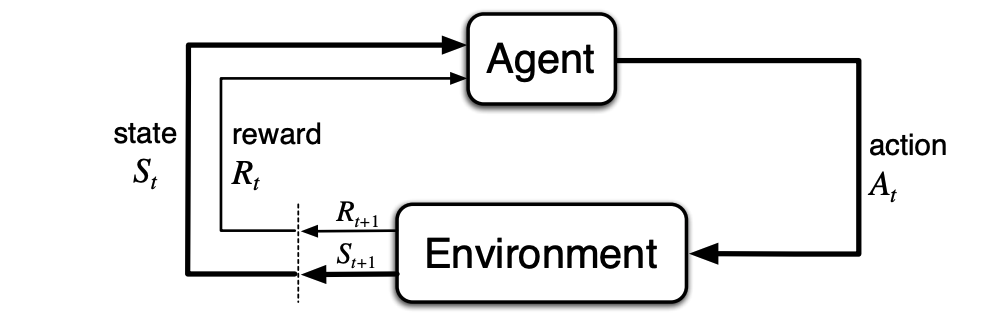}
    \caption{A reinforcement learning problem consists of an agent in an environment and states, actions and rewards \cite{Sutton2017}.}
    \label{fig:rl_suttion}
\end{figure}

At a high level, a RL problem consists of an \textit{agent} interacting with an \textit{environment}. The agent is governed by a decision-making \textit{policy}: that is, some function (possibly stochastic) which determines what the agent should do in a given state. The environment is typically described by a \textit{Markov Decision Process} with the following components: 

\begin{itemize}
  \item \textbf{States}: snapshots of the environment; observed by the agent
  \item \textbf{Actions}: means by which the agent can interact with its environment and receive a reward
  \item \textbf{Reward function}: determines (sometimes probabilistically) the reward resulting from action $a$ taken in state $s$
  \item \textbf{Transition function}: determines (sometimes probabilistically) the state that results from taking an action $a$ in state $s$
\end{itemize}
These components and workflow are summarized graphically in Figure \ref{fig:rl_suttion}. 
As the agent interacts with the environment through actions, the environment may change state, governed by the transition function. Rewards are observed that depend on both the action and the state in which it was taken. The task of an RL problem is to learn a policy $\pi(s)$ which maximises the long-run expected reward. In other words, the agent aims to maximise: 
\begin{equation}
\mathbb \sum_{t=0}^{\infty} \gamma^t r(s_t,a_t)
\label{reward}
\end{equation}

where $a_t$ is sampled from $\pi(s_t)$. Note that we have included a discount factor $\gamma$: while some problems have a termination state (known as episodic problems), others may continue indefinitely and we need to discount the rewards to prevent long-run rewards from increasing unbounded. In episodic cases, we allow $\gamma \leq 1$, otherwise $\gamma < 1$. 

\subsubsection{Solution methods}

The previous section described the general formulation of RL problems. While a comprehensive discussion of methods is beyond the scope of this white paper, some key concepts and characteristics of algorithms for solving such problems are introduced below. For a thorough treatment of RL methods, see \cite{Sutton2017}.

A large number of RL algorithms depend on the estimation of a value function. More precisely, the state-value function (typically $v_\pi(s)$) gives the expected return for following a policy $\pi$ from state $s$, while the action-value function (typically $q_{\pi}(s,a)$) gives the expected return for taking action $a$ in-state $s$ and following $\pi$ thereafter. Central to many RL algorithms is \textit{policy iteration}, where the policy is repeatedly updated to the value function, while the value function is updated to match the policy.

On the other hand, policy optimisation methods, where the policy $\pi_{\theta}(a|s)$ is some function (e.g. a neural network) which directly maps from states to actions, with parameters $\theta$. The task is, therefore, to find parameters $\theta$ giving the best policy. A variety of methods can be employed to estimate the gradient of the reward function with respect to the parameters and optimise the policy \cite{Williams1992, Sutton2000, DeBoer2005}. 

Among the principal challenges in many RL problems is the long time dependencies connecting actions and rewards. For instance, in zero-sum games such as chess and Go, the reward function is usually designed as follows: 

\[
    r(s,a)= 
\begin{cases}
    1,& \text{if agent has won}\\
    -1,& \text{if agent has lost}\\
    0,& \text{otherwise}
\end{cases}
\]

Moves early on are not rewarded until the very end of the game, leading to the well-known \textit{credit assignment problem} to correctly attribute good actions to later rewards.

\begin{figure}[h!]
    \centering
    \includegraphics[width=3in]{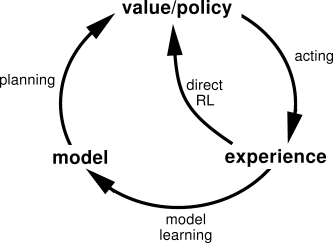}
    \caption{Model-based RL methods use estimates of the transition and reward functions to improve the policy or value functions \cite{Sutton2017}.}
    \label{fig:rl_model}
\end{figure}

Lastly, there is a noteworthy distinction between model-based and model-free methods. In model-based methods, the agent estimates the transition and reward functions, sometimes learning them from experience (Figure \ref{fig:rl_model}). Model-based methods include those used successfully in the AlphaGo algorithms \cite{SilverHuangEtAl16nature, Silver2017}, making use of tree search planning.

There is no `silver bullet' RL method which is capable of achieving expert performance for all problems. Domain knowledge is often an important contribution to developing an RL algorithm for a given problem: it is for this reason that power systems researchers are encouraged to participate in the L2RPN challenges.

\subsubsection{Formulating the Network Operation Problem}

In the electricity network context, the network operator aims to maintain a secure network state at the lowest cost. Formulation of the network operation problem depends in part on the actions and state information available to the operator, and in this section, the formulation in general terms is described. For a more detailed formulation that is relevant to this challenge, see \cite{Donnot2017}.

For RL, the network operation problem is usually described in a Markov Decision Process. In this case, the agent is the network operator, capable of performing actions on the environment, which is the power network itself. Using the MDP components described previously, a general description of the power network operation problem is given:

\begin{itemize}
    \item \textbf{States}: describes the physical network, including load, generation, line connections, time etc.
    \item \textbf{Actions}: options to reconfigure the network (topological changes) or redispatch generation.
    \item \textbf{Reward function}: depending on the problem context, measuring economic, environmental or security costs/performance.
    \item \textbf{Transition function}: determines the new state of the network following actions taken by the agent.
\end{itemize}

The operator aims to determine a policy that maximises the long-run reward in equation (\ref{reward}). Several aspects of the MDP may be customised by the operator in order to create a simpler approximation of the problem. For instance, the action space may be reduced by considering only those actions which alter the status of a single node, avoiding the action space increasing exponentially in the number of buses. Similarly, the reward function is designed by the operator: while it should generally reflect the ultimate goal of the task (in the L2RPN challenges, this is the function by which participants are scored), a hand-crafted reward function may result in better performance in practice.

Having described in general terms the formulation of the network operation problem for solving with RL, in the next section L2RPN challenges developed by this working group is outlined.

\section{Learning to Run a Power Network Challenges} 

Motivated by the pressing challenges facing network operators and the recent successes of RL, the Learning to Run a Power Network (L2RPN) challenges were originally developed by Réseau de Transport d'Électricité (RTE), the French transmission network operator, to promote investigation into the network operation problem in a competitive context. A larger working group comprising researchers from a range of institutions has sought to build on the initial successes of L2RPN to create more sophisticated, difficult and practically relevant challenges. Researchers from the machine learning and power systems communities are encouraged to develop RL agents capable of operating a simulated power network securely and at low cost. 

Given that RL methods rely on trial-and-error to improve the decision-making policy, interaction with the physical power network is not possible during training. This required the development of \textit{Grid2Op}\footnote{See: https://github.com/rte-france/Grid2Op} an open-source environment for training RL agents to operate power networks. The Grid2Op is a flexible framework, allowing researchers to accurately simulate power system dynamics for different networks while interacting with the environment through topological and redispatching actions.

\begin{figure}[t!]
     \centering
     \includegraphics[width=5in]{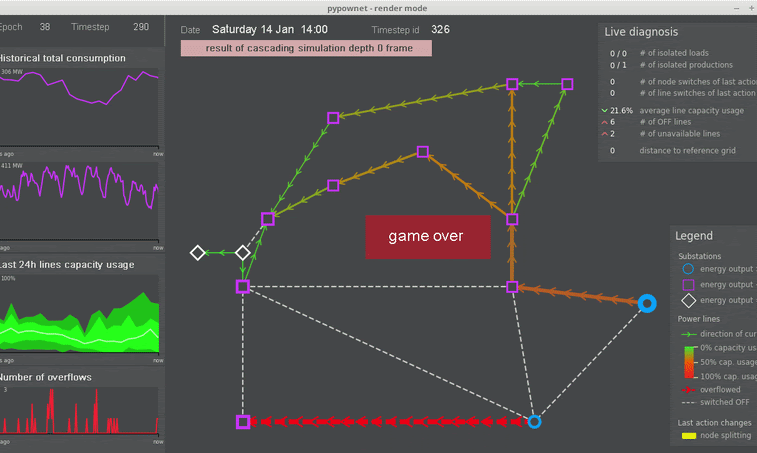}
     \caption{An illustration of the IEEE-14 simulation environment used for the L2RPN 2019 challenge.}
     \label{fig:L2RPN}
 \end{figure}

The first challenge was presented at IJCNN 2019 and was conducted on the IEEE-14 network (Figure \ref{fig:L2RPN}) \cite{Marot2019}. The goal was to operate the network under a generation and load profile by making topological changes to the network to avoid line loadings beyond their limits. The participants scored zero if operation resulted in network failure, and otherwise scored well by minimising line loadings. Several teams were successful in developing robust RL agents for network operation, with the winning team presenting a methodology based on deep Q-learning \cite{Lan2020}. 

Following the success of the 2019 challenge, further challenges will take place in 2020 with two notable developments: (1) expansion to the larger IEEE-118 network (with 118 substations instead of 14); (2) availability of redispatching actions to change the production of generators. Both of these advancements are essential for reflecting the real task of network operators. Due to the much larger state and action spaces, the 2020 L2RPN challenge is significantly harder than before, requiring novel methods to succeed. The aim of this challenge is to move further toward developing RL methods for power network operation in the real world.

\section{Conclusion}

Electricity systems around the world are changing in many ways to meet IPCC climate goals by decarbonization and by electrification of other sectors of society such as transportation and heat. As a result of these radical changes, the challenge of operating power networks in the future is becoming incrementally greater. The tools and techniques that are currently relied upon to keep the network in a secure state in real-time and to plan for the near future may no longer be adequate. 

RL is a very useful framework to develop new solution techniques to the problem of network operation in a decarbonized world. RL involves an agent performing actions, based on states of the environment and transition functions between the state and the action. By gamifying the network operation problem, RL can be deployed as a technique to develop novel solutions to network operation challenges. 
\\
\\
Following the recent success of the L2RPN 2019 challenge, in 2020 a new challenge will be launched, encompassing a larger network and the addition of redispatching actions as well as topological changes. This will be an essential step towards developing practical RL methods that can be used to assist decision-making by power network operators. This challenge aims to bring experts from the power systems and ML/RL communities together to develop solutions to solve the problems facing network operators in the decarbonised future.


\pagebreak
\section{Contributing Organizations}
The following organizations and individuals contributed to the development of the L2RPN challenge in 2020: \\ \\
\textbf{RTE:}\\
Antoine Marot - antoine.marot@rte-france.com \\
Benjamin Donnot - benjamin.donnot@rte-france.com \\
Patrick Panciatici - patrick.panciatici@rte-france.com \\
Pauline Gambier-Morel - pauline.gambier-morel@rte-france.com \\
Camilo Romero - camilo.romers@gmail.com \\
Lucas Saludjan - lucas.saludjian@rte-france.com \\\\
\textbf{L2RPN and INRIA:} \\
Isabelle Guyon guyon@chalearn.org \\
Marvin Lerousseau - marvin.lerousseau@gmail.com \\\\
\textbf{Turing Institute \& UCL:}\\
Aidan O'Sullivan - aidan.osullivan@ucl.ac.uk \\
Patrick de Mars - patrick.demars.14@ucl.ac.uk \\\\
\textbf{EPRI - Electric Power Research Institute:}\\
Adrian Kelly - akelly@epri.com \\\\
\textbf{Google Brain:}\\
Gabriel Dulac-Arnold - dulacarnold@google.com \\\\
\textbf{China State Grid - GEIRINA:} \\
Yan Zan - yan.zan@geirina.net \\
Jiajun Duan - jiajun.duan@geirina.net \\
Di Shi - di.shi@geirina.net \\
Ruisheng Diao - ruisheng.diao@geirina.net \\
Zhiwei Wang - zhiwei.wang@geirina.net \\\\
\textbf{Iowa \& Maryland Universities:} \\
Amar Ramapuram - amar@iastate.edu \\
Soumya Indela - indela.soumya@gmail.com \\\\
\textbf{TenneT:} \\
Jan Viebahn - Jan.Viebahn@tennet.eu \\\\
\textbf{Kassel University \& pandapower:} \\
Florian Schäfer - florian.schaefer@uni-kassel.de \\
Jan-Hendrik Menke - jan-hendrik.menke@uni-kassel.de \\\\
\textbf{Arizona State University:} \\
Kishan Guddanti - kguddant@asu.edu \\\\
\textbf{Encoord:} \\
Carlo Brancucci - carlo@encoord.com \\\\

 \begin{figure}[h!]
     \centering
     \includegraphics[width=3in]{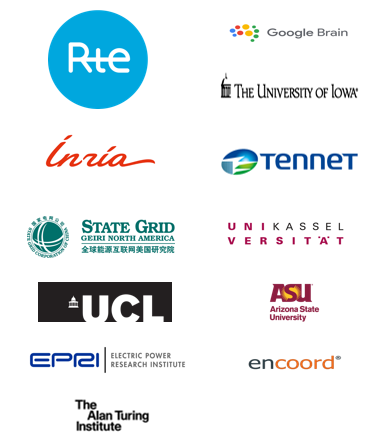}
     \label{fig:L2RPN1}
 \end{figure}

\bibliographystyle{plain}
\bibliography{references}
\end{document}